\documentstyle[12pt]{article}
\author{
Serge Galam\\ \\ \\ 
Laboratoire des Milieux D\'{e}sordonn\'{e}s et H\'{e}t\'{e}rog\`{e}nes
\footnotemark[1]\\
Tour 13 - Case 86, 4 place Jussieu, \\ 75252 Paris Cedex 05, France\\}
\title{Rational Group Decision Making \\ $\ $\\ A random field Ising model at $T=0$} 
\addtocounter{footnote}{0}
\footnotetext{Laboratoire associ\'{e} au CNRS (URA n$^{\circ}$ 800) et \`{a} l'Universit\'{e} 
P. et M. Curie - Paris 6}
\date{$\,$}
\addtocounter{footnote}{1}

\begin{document}
\maketitle
\baselineskip 3.3ex
\footskip 5ex
\parindent 2.5em
\abovedisplayskip 5ex
\belowdisplayskip 5ex
\abovedisplayshortskip 3ex
\belowdisplayshortskip 5ex
\textfloatsep 7ex
\intextsep 7ex

\begin{abstract}

A modified version of a finite random field Ising ferromagnetic model 
in an external magnetic field at zero temperature
is presented to describe group decision making. Fields may have a non-zero average.
A postulate of minimum inter-individual conflicts is assumed.
Interactions then produce a group polarization
along one very choice which is however randomly  
selected. A small external 
social pressure is shown 
to have a drastic effect on the polarization. Individual bias 
related 
to personal backgrounds, cultural values and past experiences are introduced via
quenched local competing fields.
They are shown to be instrumental 
in generating a larger spectrum of collective new choices beyond initial ones. 
In particular, compromise is found to result from the existence of individual competing bias.
Conflict is shown to weaken group polarization.  The model yields new psycho-sociological 
insights about consensus and compromise in groups.

\end{abstract}
\baselineskip 24pt
\newpage
\section{Physical approach to social behavior}

Some attempts to use  Statistical Physics to decribe various aspects of social behavior
started long ago [1], for instance to study strike process [2]. They 
are getting
more numerous in recent years. Among others, we can cite  
political organisation [3], group power dynamics [4], social impact [5], outbreak of cooperation [6],
stock market [7], the DNA analyse [8], evolution theory [9] and ageing [10].
	
Social systems often involve cooperative behavior of 
some small or large numbers of people. The main difficulties in 
studying these systems is believed to result from the rich variety of existing individual 
features [11, 12]. The
 complexity of a group could thus be expected to increase with its size. 
However crowds, which contain large numbers of persons, behave in some aspects 
 like one collective individual making some behavior even simpler than in the case of one 
 individual [13].
 
The theory of critical phenomena can indeed shed light on above paradox.
In particular, the two basic concepts of universality and 
 irrelevant variables [14] which mean physical characteristics like the 
form of microscopic interactions and their
 physical nature have no effect on the universal properties of
the collective behavior, are of importance to tackle social behavior. 

Along these two concepts, it makes sense to suppose there exists in social systems too,
on the one hand, many properties 
 associated with purely individual characteristics, and on the other, some few properties 
 which produce the collective social state [15]. 
 More precisely, the hypothesis behind the present approach is that micro-macro relationships
 are universal and hold true beyond the nature of the various entities involved. 

In this paper we study group decisison making, in particular conditions
 which lead to either a polarization or a compromise of the group [15, 16]. 
Here ``polarization" means that most of the people move 
in one direction [17].
 	
To keep the presentation simple, we used a model in which a group of N persons has to make
 a decision restricted to two options. Each 
 person can choose between only  ``yes" and ``no".
 The model is articulated around a {\it Postulate} of minimum pair individual conflict.
 Competing interactions are also introduced via local quenched biases.

Formally we are using a modified version of the random field Ising ferromagnetic model 
in an external magnetic field at zero temperature. However here the system is finite in size. 
Moreover ``random fields" may have a non-zero configurational average. 
Results may depend on the field configuration.

Our model does not aim to novelty in Statistical Physics but instead
sheds a new light on various aspects of human behavior.  Moreover it provides ground to 
a class of possible social experiments. Thus we tried to write the paper such 
that it can be understood by non-physicists.
 
Section 2 considers the simplest situation introducing
the {\it symmetrical individual}. It is then extended to the {\it symmetrical group} in section 3.
Interactions are introduced as well as a measure of the group conflict.
The concept of a {\it symmetry breaking choice} is defined in Section 4. 
Section 5 deals with the group emergence from isolated individuals
to integrated group members.
The {\it group formation process} is analysed using individual anticipating. 
In particular we study mechanisms by which 
 either a compromise or a polarization of the group is produced.
Social 
 surrounding pressure and individual differences are included in Section 6 with the {\it bias individual}.
This bias accounts for individual representations [18] which are well established in social sciences. It 
results from cultural values, 
 beliefs and personal
 experiences.
The model is illustrated through a few examples in
Section 7. We present respectively cases of balanced equal representations, balanced unequal 
representations,
minorities and leaders. The last Section contains concluding remarks about extension of the model to non-rational 
behavior, i. e., non-zero temperature.


\section{The symmetric individual}
We start from the simplest situation in which one 
individual has to make a choice between two
answers either yes or no. 

Such cases are indeed numerous 
in the social world. Moreover cases with a larger spectrum of answers can usually be mapped
at some approximative level into a two-answer case [17]. The 
individual choice can then be 
represented 
by a two-valued variable $c$ with $c=1$ associated to answer yes and $c=-1$ to answer no.

For a collection of $N$ persons  each individual choice is 
represented by a
variable $c_i$,
where $i=1,2,...,N$ and $c_i=\pm 1$. 

A collective choice of 
the  $N$ person group can be defined as the simple sum of each individual choice,
\begin{equation}
C=\sum_{i=1}^{N}c_i\:.
\end{equation}
From Eq. (1) it is seen that aggregation enlarges drastically the spectrum of
possible choices. It actually increases from $2$ at the individual level up to $2^N$
at the group level. Nevertheless to materialize this spectrum 
a structure to collect individual
answers, to sum them up and to display the net result is necessary. 

Moreover, to go beyond initial 
twofold answer requires some complex internal transformation in order to associate a meaning 
to each one
of the $2^N$ answers. However
the use of some rules, for instance a majority rule, can bring the collective choice back 
to the individual one with only two answers.

\section{The symmetrical group}

At this stage yes and no are equiprobable.
 The probability distribution function is therefore,
\begin{equation}
p(c_i)=\frac{1}{2}\{ \delta (c_i-1)+\delta (c_i+1)\}\:,
\end{equation}
where $\delta$ is the Kronecker function.

Considering $N$ isolated individuals, each one makes its choice without any 
interaction with others. 
The probability distribution for the collection of $N$ persons is thus,
\begin{equation}
p(C)=\prod_{i=1}^{N}p(c_i)\:,
\end{equation}
where $p(c_i)$ is given by Eq. (2).

From Eqs. (2) and (3) the collective choice $C$ is indeed zero on 
average, with fluctuations of order $\frac{1}{\sqrt{N}}$. The result $C=0$ 
creates a new qualitative choice which did not exist at the individual level.

$C=0$ can be understood as the perfect compromise choice. Since the macroscopic quantity $C$
is zero, the aggregation process turns out to have
no effect at the macro-level making the associated group symmetrical. Perfect
compromise means no group existence as such, it is a a neutral group with no link 
among group members. 

We now proceed to introduce interactions. Given a pair of persons
$i$ and $j$, only four different choice
configurations can be produced. These are,
(1) $c_i=c_j=+1$,
(2), $c_i=c_j=-1$,
(3), $c_i=-c_j=+1$,
and (4), $c_i=-c_j=-1$.
In configurations 1 and 2 the two persons $i$ and $j$ are making the same choice.
They disagree in configurations 3 and 4
with opposite choices, they are at conflict.
However agreement or conflict materializes only if $i$ and $j$ are both aware of
the other's choice, in other words, only once they are somehow interacting. 

We can thus naturally identify a state of conflict (configurations 3 and 4) or agreement
(configurations 1 and 2) using the product $c_ic_j$. It
is equal to $+1$ in agreement and to $-1$ in conflict. From now on, agreement is defined as a positive 
conflict. Both cases do
 not differentiate which choice is actually made, in accordance with the group symmetrical nature.

However prior to the decision itself, both individuals may, for instance, argue for a long 
time, or
exchange written information. On the other hand, they may decide without any discussion.
We thus need to introduce a quantity to measure this choice involment.
Let us call $I$ the exchange amplitude. This parameter can be incorporated
into the configurational choice labelling using the product $Ic_ic_j$ instead of $c_ic_j$.
 An agreement state is associated with ($+I$) while 
($-I$) corresponds to a conflict state. $I$ measures the conflict amplitude. 

Restricting interaction to pairs the overall group
conflict is measured by the function, 
\begin{equation}
G_I \equiv I\sum_{<i,j>}c_ic_j\:,
\end{equation}
where we have assumed that the exchange amplitude $I$ is constant for all interacting $(i,j)$
pairs. We call $G_I$ the group internal conflict function and $<i,j>$ represents all 
interacting pairs. 

\section{The symmetry breaking choice}

The internal conflict function $G_I$ measures the conflict amplitude in a group for each one of
the $2^N$ decisional configurations. It discriminates among various possible choices, 
but does not 
indicate which one is chosen by the group. For the group decision dynamics to operate, it is
necessary to 
invoke a criterion to select which among possible states is favored by the group. 

Along the minimum ernergy principle, we introduce a {\it Postulate} to determine the group
dynamics direction. It reads,
\begin{center}
{\it ``Each individual favors the choice which minimizes its own conflict".}
\end{center}
A given person will thus select its choice according to a minimum conflict principle. 
Justification of this {\it Postulate} 
is beyond the scope of the present work. It will be motivated a postiori by the results obtained 
from the model. Minimum conflict means maximum agreement. 

To grasp part of the decision making dynamics we can start randomly from one person who made at random
a choice of either yes or no. Afterwards it does not change its choice. 
Then all persons interacting with this 
particular person will make the same choice, to minimize their own conflicts. Within a sequence process
all people interacting 
with them will again make the same choice to favor agreement.
In so
doing the whole group will end up making the same initial choice the first person did select.

The net result of these dynamics is an
extreme polarization of collective choice with
$C=\pm N$. The sign, i. e., the polarization direction,
is determined by the initial individual choice which was made randomly.
This polarization
phenomenon holds whoever is chosen to be the initial person. Only the direction (yes or no) will change
from one initial person to another one.

In real life situations, the above process starts simultaneously from several persons.
The dynamics of choice spreading is a rather complex phenomenon. Monte Carlo simulations 
on zero-temperature dynamics on the Ising model showed indeed
non trivial behavior at all dimensions [19].  However, the group succeeds somehow to select 
only one intial choice at long distance
allowing everyone to minimze its own conflicts.
Therefore we conclude, 
\begin{center}
{\it Symmetrical groups polarize themselves towards an extreme choice. 
The direction
of that choice however is arbitrary. Each extreme is equiprobable.}
\end{center}

From the definition of the group internal conflict function $G_I$ (Eq. 4) and 
above polarization 
result,
the {\it  Postulate } of
individual minimum conflict turns out to be identical to maximize $G_I$.

The polarization effect which results from group member interactions is identical to the
{\it Spontaneous Symmetry Breaking} phenomenon well known in the physics of collective phenomena [14]. 
Individual local interactions  make the group to behave as one {\it super-person} 
[13]. That {\it super-person} 
chooses between two possible choices with equiprobability likewise the isolated individual. 

In
parallel the individual within the group has lost its
freedom of choice. It must now make a choice identical to people it interacts with. Individual
freedom has been given up in favor to group freedom. 

Here perfect compromise has 
disappeared. Simultaneously  the group
decision produces an effect on its surrounding somehow
proportional to $N$ since $C=\pm N$. Without interactions 
 $c_i=\pm 1$ individual 
were overall self-neutralised macroscopically. Interactions have produced strong individual 
correlations associated with a {\it  Symmetry Breaking}.

Our finding sheds new light on results obtained from group decision making experiments conducted 
in social psychology. The polarization effect was clearly evident in data reported in [17].
 However, until now, most theoretical explanations have been 
unconvincing in connecting choices at respectively the individual level 
[12] and the group level [16]. 
Our proposal is that polarization effect arises quite naturally from first principles, i. e., from interactions.

\section{Anticipating effect}

At this stage we need 
to formalize the internal group dynamics which proceeds from initial 
individual choices towards the final collective choice. The exchange term must be modified
to account for the emergent group reality. We first rewrite $G_I$ as,
\begin{equation}
G_I=\frac{I}{2}\sum_{i=1}^{N}\left\{\sum_{k=1}^{n}c_{j(k)}\right\}c_i\:.
\end{equation}
where $n$ is the number of persons one individual interacts with. To keep the presentation simple
this number is assumed equal for
everyone. In case  everyone interacts with everyone $n=N$.

Now we modify Eq. (5) to account for the process
of group formation. People do anticipate the emergence of a collective choice. Each
individual $i$ will thus try to project through its partner's choices $c_j$ (the people $i$ discusses
with), its expectation of the overall final group decision. 

Individual $i$ then extrapolates the $j$'s choice
$c_j$ to the expected collective choice the group will eventually make without its own
participation. Within this process, individual $i$ perceives the $j$'s choice as given 
by the transformation,
\begin{equation}
c_j \rightarrow \frac{1}{N-1}(C-c_i)\:,
\end{equation}
where $C$ is the collective choice defined as before.
Once this process is completed, Eq. (5) becomes,
\begin{equation}
G_I^g=\frac{I}{2}\sum_{i=1}^{N}\left\{\sum_{j=1}^{n}\frac{1}{N-1}(C-c_i)\right\}c_i\:,
\end{equation}
and,
\begin{equation}
G_I^g=\frac{In}{2(N-1)}\left\{C\sum_{i=1}^{N}c_i-\sum_{i=1}^{N}c_i^2\right\}\:,
\end{equation}
where superscript $g$ denotes active anticipating process.
Using the collective choice definition $C=\sum_{i=1}^{N}c_i$ and the property $c_i^2=1$ we get,
\begin{equation}
G_I^g=\gamma  \frac{C}{N}\sum_{i=1}^{N}c_i-\gamma  \:,
\end{equation}
where $\gamma  \equiv\frac{nIN}{2(N-1)}$ is a constant independent of the group choice. As such 
it is irrelevant to the collective choice. $C$ is not yet the
final decision. Rather it is the expected final collective choice.
We can rewrite Eq. (9) in the form,
\begin{equation}
G_I^g=S_g\sum_{i=1}^{N}c_i-\gamma  \:.
\end{equation}
where
\begin{equation}
S_g \equiv\gamma  \frac{C}{N}\:
\end{equation}
acts as a group field which couples with each individual choice. The field notion is a
natural way to account for some pressure towards a definite choice.  
Within our convention of minimum conflict  the  
product $S_gc_i$ measures that influence. 
A positive field  $S_g$ favors
a positive choice $+1$, while $-1$ is associated to a negative field. The conflict 
amplitude is given by  $S_g$.

We have indeed a self-consistent expression
since on one hand, individual $i$ wants to go along the virtual field $S_g$, and on the other
hand it contributes directly to this virtual field through its dependance on the collective 
choice $C$.
Rewriting Eq. (10) as 
\begin{equation}
G_I^g=\gamma  \frac{C^2}{N}-\gamma  \:
\end{equation}
shows that maximizing $G_I^g$ results in 
maximizing $C^2$ which is obtained by $C^2=N^2$. It is an extreme polarization with
either $C=+ N$ or $C=- N$.
Again individual minimum conflicts  appear clearly identical to 
the maximum of the group internal conflict function.

At this stage of the model our ``group formation process" is formally
identical to the mean 
field theory of phase transitions. While in physics, it is an approximation, here it embodies the social 
mechanism of anticipation.

\section{The bias individual}

We are now in position to overcome two simplifying assumptions made earlier. First,  
most choices an individual and a group have to make are not independent of 
the surroundings  as assumed above. We now will account for pressure applied to the group from 
the outside. Second, assuming identical individuals, with no apriori individual differences
in preferences
about the issue, does not hold in most cases. Individual differences 
will now be included.

\subsection{Social pressure}

The existence of external pressure on group members means  
the equiprobability hypothesis (Eq. 2) no longer holds.
It is achieved introducing a 
quantity which differentiates the two possible choices. We call this quantity the social 
field $S$. It turns the symmetrical individual into 
a social one with now  $p\:(c_i=1)\neq p\:(c_i=-1)$. 

Similarly with above group field $S_g$, 
each person's conflict with $S$ is represented
by the product $Sc_i$. Agreement is associated with $Sc_i>0$, i. e., the choice is made along
the field with $S$ and $c_i$ having the same sign. In contrast $Sc_i<0$ represents a 
conflict between the individual and the social pressure, since $S$ and $c_i$ have opposite signs.
The surrounding group conflict measure is,
\begin{equation}
G_S\equiv \sum_{i=1}^{N}Sc_i\:.
\end{equation}
Applying the {\it Postulate} to the sum $G_I^g+G_S$ still results in an extreme polarization
but now its direction is no longer random.
The group choice is $C=+N$ for $S>0$, and $C=-N$ with $S<0$. Under external pressure,
even extremely weak, the 
group and the individual behave identically. They both follow the pressure 
induced by the external 
pressure. The
{\it Symmetry Breaking} choice is no longer random. 

Here the {\it super-person} represented by the whole group is identical to 
the individual person. They are both aligned along the field. This result is at contrast with
the symmetrical state, where 
the individual loses its freedom of choice in favor of the group choice freedom. 

\subsection{The Representational  state}

To get closer to experimental situations, individual differences in preferences about some issues
 are now introduced. Following social literature they are called ``individual representations" [18].
A representation 
varies
in both, direction and amplitude, from one person to another. It depends  upon cultural 
values, past experiences, ethics and beliefs. It is attached to a person.

The representational effect can
be included within our formalism by introducing an additional field. We call $S_i$ the
internal social field attached to individual $i$.
 Its properties are similar to those of a social field $S$. The difference
being that the social field applies uniformly to each group member while an internal social field
acts only on one person. 

As for the other fields, conflict with the
internal social field is accounted for in
the product $S_ic_i$. It is positive for a choice made along 
the representation  (internal 
agreement with personal values), and negative otherwise (internal conflict with
personal values).
The group representation conflict  measure is given by,
\begin{equation}
G_R\equiv \sum_{i=1}^{N}S_ic_i\:.
\end{equation}

The distribution of individual representations  is required to determined the group collective choice.
 The representation  effect is enhanced in the isolated-person
 case where both exchange amplitude and social external field are zero.
There, from the {\it Postulate}, final decision is found to result from every individual 
following its own representation. 
It gives, 
\begin{equation}
C=\sum_{i=1}^{N}\frac{S_i}{|S_i|}\:,
\end{equation}
where the $|...|$ denotes absolute value.

This equation illustrates the qualitatitve change driven by the  existence of representations.
Actual $C$ value 
can now vary over the whole 
spectrum of values $-N,...,0,...,+N$. Compromise $C=0$ can again be an outcome. 
Individual representations  are thus instrumental for making the whole model 
relevant to real situations in which collective choices are far richer than $C=\pm N$.

In others words, prior to group formation, individuals have their own representations 
which determine their 
apriori answers to the initial question. All these representations
result 
in either yes or no. Then, in the process of group formation, people start to
interact through the  yes and no distribution in the group. 

However to reach a collective 
choice, due to the existence of opposite representations, people must construct new answers in 
addition to the initial yes and no. Answers are thus enriched 
during
group formation, due to driving representations. On the other hand, within the 
neutral state groups do not produce new answers. 

Once the final decision is reached, each group member identifies 
with the collective choice triggering its new individual choice to $d=\frac{C}{N}$ which may differ 
from the initial
$c_i$. In the neutral state $d=\pm 1$.

Thus, in the process of building a new answer $d$, a new representation  has been 
produced by the group. This new representation 
is integrated by each individual to yield the $d$ choice. Group formation has 
qualitatively modified 
individual representations. 

This process shows that while individuals resist adopting a  
representation opposed to their own, via the group transformation, they will join a new
common representation which accounts for the overall balance of initial representations. 
The preceding takes place around a new answer which did not exist prior to the group forming.

Note our qualitative departure from usual Statistical Physics. Here we are not considering 
an average individual position, but a well defined and fixed individual position. This position
results from 
the group forming. In most cases $d$ is different from $\pm 1$. We are thus passing from a class of Ising 
variables $c_i=\pm 1$ to one continuous variable $\frac{-1}{N}\leq d \geq \frac{+1}{N}$.

\subsection{The frustrated individual}

Adding together
all the effects introduced until now results in an extended group internal conflict function
$G=G_I^g+G_S+G_R$ which is,
\begin{equation}
G=I\sum_{i,j}c_ic_j+S\sum_{i=1}^{N}c_i+\sum_{i=1}^{N}S_ic_i\:.
\end{equation}

The extended form of Eq. 
(16) makes maximizing $G$ a more difficult task since competing effects are active. 
A given individual wants now to minimize its overall own 
conflict. There exist three contributions,
\begin{itemize}
\item Interacting group members: the individual wants to come up with the same final decision as
preferred by interaction
partners.
\item External social field: the individual wants to comply to the external pressure from 
immediate surroundings.
\item Internal social field: the individual wants to comply to the internal pressure from its
personal representations.
\end{itemize}
These three elements are not necessarily satisfied simultaneously. From the
{\it Postulate}, the individual wants to minimize overall personal conflict. It could result in 
simultaneous agreement with some of above items, and conflict with others. It
is clearer to write Eq. (16) as,
\begin{equation}
G=\sum_{i=1}^{N}S_{g,i}^rc_i-\gamma  \:,
\end{equation}
where,
\begin{equation}
S_{g,i}^r=S_g+S+S_i\:,
\end{equation}
is the resulting field applied to individual $i$ in the group formation process.
Maximum $G$ and minimum individual conflicts are achieved when
each individual follows his
resulting field sign. If $S_{g,i}^r>0$, then $c_i=1$ and $c_i=-1$ for  $S_{g,i}^r<0$. 

The case
$S_{g,i}^r=0$ results in an undetermination of the $i$ choice as in the isolated
neutral case. Satisfying $S_{g,i}^r$ sign does imply satisfying
simultaneously $S,\:S_i$, and $S_g$ signs. This competing effect is the signature of the  
psychological complexity involved in the decision making process. 
Each person first follows its resulting field $S_{g,i}^r$ to produce a collective choice $C$.
Then this collective choice is individualy integrated back with  $c_i\rightarrow 
d=\frac{C}{N}$.

\section{Illustration of the model}

To gain a deeper insight about the meaning of Eqs. (17) and (18), we now analyse four different
specific cases. Each will illustrate some basic feature of the model.

\subsection{Two balanced opposite biases case}

We consider an evenly divided group of $N$ persons with no external social field, i. e.,
 $S=0$. Half
the persons have a positive representation $S_i=+S_0$, and the other half have a negative 
representation with
the same amplitude $S_j=-S_0$. The whole group has thus no net representation. Interactions are of
amplitude $I$ and each person discusses with $n$ other persons. In small, face-to-face groups,
everyone usually interacts with everyone else, so $n=N$.
The corresponding internal conflict function is,
\begin{equation}
G=-\gamma  +\frac{\gamma  }{N}C^2+\frac{N}{2}(S_0c_i^+-S_0c_j^-)\:,
\end{equation}
where $c_i^+$ and $c_j^-$ are attached to 
persons with respectively positive and negative representation. The constant 
$\gamma  \equiv\frac{nIN}{2(N-1)}$ 
has been introduced earlier in the group formation section. The collective choice may be written as
$C=\frac{N}{2}(c_i^++c_j^-)$.
The actual choice is the one which maximises $G$. In this case it is easily singled out, since 
there
exist only two different kinds of persons symbolised by $c_i^+$ and $c_j^-$. Four choice
configurations are possible,
(a) $c_i^+=+1;\: c_j^-=+1;\:C=N$, (b) $c_i^+=-1;\: c_j^-=-1;\:C=-N$, (c) $c_i^+=+1;\: c_j^-=-1;\:C=0$,
(d) $c_i^+=-1;\: c_j^-=+1;\:C=0$.

The first two (a and b) are agreement and others (c and d) are conflict. Associated internal
conflict functions  are, $G(a)=G(b)=-\gamma  +N\gamma$, $G(c)=-\gamma  +NS_0$,
$G(d)=-\gamma  -NS_0$.

Clearly $G(d)<G(c)$, reducing the choice to either (a and b) or (c). In case 
$\gamma  >S_0$, we have $G(a, b)>G(c)$, indicating that the interaction strength  
proportional to $nI$ is stronger than $S_0$. The group then polarizes with $C=\pm N$.
The
direction of the extreme choice occurs at random. 

Half of the members are fully 
satisfied with both their 
representation and their partners while the other half is in conflict with its own representation.
This result means in particular that the ``losing" subgroup has to build a new representation 
which embodies
some level of internal conflict. The ``winning" part does not modify its initial representation.
In this case, no new answer was built. We have $c_i\rightarrow d=\pm 1$.

On the other hand, strong representation, i. e., $\gamma  <S_0$ favors compromise, with 
the collective 
choice $C=0$. Each member $i$ starts from a personal representation to decide eventually through 
weak interactions on a medium compromise, with the creation of a new answer $d=0$.  
Again, this compromise choice
did not exist prior to the group formation. It is the result of cooperation between the 
group level 
and the individual level.
\begin{center}
{\it Within a balanced representation group, exchange favors a compromise. Weak exchange
result in an extreme polarization along a random direction.}
\end{center}

\subsection{Two unbalanced opposite biases case}

We now go back to the previous example, but consider a stronger positive representation. This
is done by writing 
the negative representation fields as $S_j=-\alpha S_0$, with $0<\alpha <1$. Respective 
numbers of positive and 
negative representations  are equal. Only the internal conflict function
values are changed to become respectively,  
$G(a)=-\gamma  +N\gamma   +\frac{N}{2}(1-\alpha)S_0$,
$G(b)=-\gamma  +N\gamma   -\frac{N}{2}(1-\alpha)S_0$,
$G(c)=-\gamma  +\frac{N}{2}(1+\alpha)S_0$,
$G(d)=-\gamma  -\frac{N}{2}(1+\alpha)S_0$. 

Since $0<\alpha <1$, $G(a)<G(b)$ and  $G(d)<G(c)$, always. However in the
case $\gamma  >S_0$ the polarization direction is determined with $C=+N$. 

Before we had $\alpha=1$ which
made the direction arbitrary, but now it is the strongest initial representation which wins. 
The discussion
process within the forming
group has made the weaker-biased people align themselves with the stronger ones. Here we have 
$c_i\rightarrow d= 1$. 

In order for a
compromise outcome to be favored, a decrease in exchanges among group members is required. 
For $\gamma  <S_0$, the
final choice is $C=0$ which gives $c_i\rightarrow d=0$. 
\begin{center}
{\it Within an unbalanced representation group, exchange favors the initially strongest
representation.
Only a limitation of exchange may produce a compromise.}
\end{center}
 
\subsection{The minority case}

Most cases do not have an equal number of people in two opposite subgroups.
Usually there exist a majority and a minority. Let us consider a minority number M of people 
$(M<\frac{N}{2})$ with a positive representation $S_i=+S_0$. The majority then contains  $(N-M)$
with an unequal 
negative representation $S_j=-\alpha S_0$. We chose, for instance, the case of a minority 
more motivated  
than the majority, i. e., $0<\alpha <1$. 

Denoting $c_i^+$ and $c_j^-$ the respective
minority and majority choices, the collective choice is given by $C=Mc_i^++(N-M)c_j^-$.
The internal conflict function is $G=\frac{n}{2(N-1)}IC^2+(Mc_i^+-(N-M)\alpha c_j^-)S_0$.
Associated four choice configurations are,
(a) $c_i^+=+1;\: c_j^-=+1;\:C=N$,
(b) $c_i^+=-1;\: c_j^-=-1;\:C=-N$,
(c) $c_i^+=+1;\: c_j^-=-1;\:C=2M-N$,
(d) $c_i^+=-1;\: c_j^-=+1;\:C=-2M+N$. 

The first two (a and b) are minority-majority agreement while the others (c and d) are
minority-majority
conflict. Associated internal conflict functions  are,  
$G(a)=-\gamma  +N\gamma   +\{(1+\alpha )M-\alpha N\}S_0$,
$G(b)=-\gamma  +N\gamma   -\{(1+\alpha )M-\alpha N\}S_0$,
$G(c)=-\gamma  +\frac{\gamma  }{N}(2M-N)^2+\{(1-\alpha )M+\alpha N\}S_0$
$G(d)=-\gamma  +\frac{\gamma  }{N}(2M-N)^2-\{(1-\alpha )M+\alpha N\}S_0$. 

Analysis of above expressions is complicated since now several parameters are involved. There are
$N$, $M$, $nI$, $S_0$ and $\alpha$. Let us comment on some cases. Again, case (d) is never selected
since indeed nothing is satisfied in that case, neither interactions nor representations.
\begin{itemize}
\item Interaction effects are winning over representation effects: the group is polarized, i. e., 
$G(a)$ or $G(b)>G(c)$.
\begin{description}
\item{(a)}
The  minority wins, turning the majority to its side if $G(a)>G(b)$. It is the case 
if $M<(N-M)\alpha$. Condition $G(a)>G(c)$ is ensured by 
$nIM>(N-1)\alpha S_0$.
\item{(b)}
The majority wins, turning the minority to its side if $G(a)<G(b)$. It is the case if 
 $M>(N-M)\alpha$. Condition $G(b)>G(c)$ is ensured by $nI(N-M)>(N-1)S_0$.
The condition does not depend on $\alpha$ since in both cases (b) and (c) the majority follows
its own representation $-\alpha S_0$.
\end{description}
\item Representation effects are winning over interaction effects: the group is balanced, i. e.,
$G(a)$ and $G(b)<G(c)$.

Condition $G(a)<G(c)$ is ensured by $nIM<(N-1)\alpha S_0$.
Condition $G(b)<G(c)$ is ensured by $nI(N-M)<(N-1)S_0$.
A balanced collective choice reflecting the respective strength in numbers of each group
is given by case (c). 
\end{itemize}

\subsection{The leader case}

In most groups, persons are not all equal in status. The inequality can stem from either a 
strong character
or an institutional position, like for instance a group president who has a tie-breaking vote. To 
account for such 
situations it is enough to associate a stronger individual field to the leading person in the group.
In other words the leader case is a special case of a strong minority which reduces
to one person.

We can thus use the minority case equations putting $M=1$. In the case
of a charismatic leader we take $\alpha \sim 0$ 
with $0<\alpha<1$ to emphasize its weak aspect. However for another kind of leader, 
the authoritarian for
instance, an external field would account for the pressure the leader applies 
to all group 
members.

Let us consider a leader with a positive representation $S_1=+S_0$. The majority figure is 
then $N-1$ with 
an unequal 
negative representation $S_j=-\alpha S_0$. Denoting $c_0^+$ and $c_j^-$ as the respective
leader and majority choices. The collective choice is given by $C=c_0^++(N-1)c_j^-$.
The internal conflict function is $G=\frac{n}{2(N-1)}IC^2+(c_0^+-(N-1)\alpha c_j^-)S_0$.
The associated four choice configurations are,
(a) $c_0^+=+1;\: c_j^-=+1;\:C=N$,
(b) $c_0^+=-1;\: c_j^-=-1;\:C=-N$,
(c) $c_0^+=+1;\: c_j^-=-1;\:C=2-N$,
(d) $c_0^+=-1;\: c_j^-=+1;\:C=-2+N$.

The first two (a and b) are leader-majority agreement while the others (c and d) are 
leader-majority
conflict. Associated internal conflict functions  are,
$G(a)=-\gamma  +N\gamma   +\{(1+\alpha )-\alpha N\}S_0$,
$G(b)=-\gamma  +N\gamma   -\{(1+\alpha )-\alpha N\}S_0$,
$G(c)=-\gamma  +\frac{\gamma  }{N}(2-N)^2+\{(1-\alpha )+\alpha N\}S_0$,
$G(d)=-\gamma  +\frac{\gamma  }{N}(2-N)^2-\{(1-\alpha )+\alpha N\}S_0$. 

Analysis of above expressions is as complicated as in the minority case  with parameters
$N$, $nI$, $S_0$ and $\alpha$. Let us comment on some cases. Again, case (d) is never selected
since indeed nothing is satisfied in that case, neither interactions nor representations.
\begin{itemize}
\item Interaction effects are winning over representation effects: the group is polarized, i. e.,
$G(a)$ or $G(b)>G(c)$.
\begin{description}
\item{a}
The  leader wins, turning the majority to its side if $G(a)>G(b)$. It is the case 
if $\frac{1}{\alpha}<N-1$. Condition $G(a)>G(c)$ is ensured by $nI>(N-1)\alpha S_0$.
\item{b}
The majority wins, turning the minority to its side if $G(a)<G(b)$. It is the case if 
$\frac{1}{\alpha}<N-1$. Condition $G(b)>G(c)$ is ensured by $nI>S_0$.
The condition does not depend on $\alpha$ since in both cases (b) and (c) the majority follows
its own representation $-\alpha S_0$.
\end{description}
\item Representation effects  are winning over interaction effects: the group is balanced, i. e., 
$G(a)$ and $G(b)<G(c)$.

Condition $G(a)<G(c)$ is ensured by $nI<(N-1)\alpha S_0$ and
condition $G(b)<G(c)$ by $nI<S_0$.
A balanced collective choice reflecting the respective numerical strength of each group
is given by case (c). 
\end{itemize}


\section{Conclusion}

A simple Ising-like model has been presented to describe group decision making. 
It is indeed a modified version of the random 
field Ising 
ferromagnetic model 
in an external magnetic field at zero temperature. however our system is finite in size
and fields may have a non-zero configurational average. 
in principle results may also
depend on the field configuration. Moreover we crossover in the group decision making process
from a class of Ising variables to one continuous variable.

The hypothesis behind our approach is that  group decision making obeys universal laws which are
independent of the nature of the issue at stake.
Our main results with respect to the qualitative properties of group decision making are,
\begin{description}
\item[*] Exchanges among individual does not aim to select an issue, but rather to align people 
along the same issue. The issue itself is random with respect to exchanges.
\item[*] Exchanges among individual does not favor compromise about an issue. On the opposite it 
produces polarization, i. e., extreme options.
\item[*] Reducing exchanges favors compromise. 
\item[*] External social pressure is extremely efficient on selecting an option.
\item[*] Individual bias is a necessary ingredient to both weaken extreme option and oppose an
external social pressure.
\end{description}

These theoretical results must be put in parallel to various data obtained from a large number 
of experimental
studies which show groups polarize along
an extreme position reflecting the dominant pole of attitudes and not around an average position as 
a priori expected [16, 17]. 

Our emphasize is on building a conceptual methodology rather than a final complete 
theory.
In a forthcoming paper we will introduce non-rational behavior which is a real life 
basic feature. It will be analogus to temperature. However within our model we will define 
a ``local temperature" in a finite system.

\subsection*{Acknowledgments}

I would like to thank Dietrich Stauffer for stimulating comments on the manuscript.
 

\end{document}